\documentclass[twocolumn,pre,superscriptaddress]{revtex4-2}
\usepackage{float}
\usepackage{graphicx}
\usepackage{tikz}
\usetikzlibrary{shadows.blur}
\usetikzlibrary{arrows,automata,snakes}
\usetikzlibrary{angles,
	quotes}
\usepackage{tikz-qtree}

\usepackage{amssymb,amsmath}   
\usepackage{dsfont} 							

\usepackage{bm}
\usepackage{dcolumn}

\usepackage{verbatim}

\usepackage{hyperref}
\hypersetup{backref,colorlinks=true,breaklinks,urlcolor=blue,linkcolor=blue,citecolor=blue}

\usepackage[capitalise]{cleveref}
\usepackage[notrig]{physics}   
\usepackage{braket}

\renewcommand{\hat}[1]{#1}    
\renewcommand{\rho}{\varrho} 
\renewcommand{\i}{\mathrm i}  

\begin{document}

\title{Optimal form of time-local non-Lindblad master equations}
\author{Tobias Becker}
\email{tobias.becker@tu-berlin.de}
\affiliation{Institut f\"ur Theoretische Physik, Technische Universit\"at Berlin, Hardenbergstrasse 36, 10623 Berlin, Germany}
\author{Andr\'{e} Eckardt}
\email{eckardt@tu-berlin.de}
\affiliation{Institut f\"ur Theoretische Physik, Technische Universit\"at Berlin, Hardenbergstrasse 36, 10623 Berlin, Germany}

\begin{abstract}
Time-local quantum master equations that describe open quantum systems beyond the limit of ultraweak system-bath coupling are often not of Gorini-Kossakowski-Sudarshan-Lindblad (GKSL) form. 
Prominent examples are the Redfield equation approximating general open quantum systems and the Hu-Paz-Zhang equation exactly describing a damped harmonic oscillator. 
Here, we show that not only the former, but also the latter can be brought to pseudo-Lindblad form, with a dissipator that resembles that of a GKSL equation, except for the fact that some of the terms have negative weights. 
Moreover, we systematically investigate transformations that leave the dissipator of pseudo-Lindblad equations unchanged, while changing the relative weight between its positive and negative terms. 
These can be used to minimize the weights of the negative terms, which is optimal both for the convergence of a recently developed quantum-trajectory unraveling of pseudo-Lindblad equations as well as for the truncation of the negative terms to obtain a GKSL equation. 
\end{abstract}

\maketitle
\section{Introduction}
In many circumstances the coupling of a quantum system to its environment matters and has to be taken into account for predicting its transient or steady state. 
At the same time, we are often not able to fully describe the state of the environment. 
Unless the system is in thermal equilibrium, and thus captured by statistical mechanics \cite{campisiFluctuationTheoremArbitrary2009,TalknerHaenggi2020,Alicki22,TimofeevTrushechkin2022,Anders2022}, we, therefore, need to resort to effective descriptions of the system that take the impact of the environment into account at least on an approximate level.
This can be done with the help of quantum master equations for the reduced density operator of the system \cite{FBloch56,AGRedfield65,GoriniKossakowski1976,GLindblad1976,RadiationConsideredReservoir1998,CWGardiner00,breuerpetruccione,wallsQuantumOptics2008,HuPazZhang1992,KarrleinGrabert97,Paz94,GrabertWeiss84}. 
For ultraweak system-bath coupling (being small compared to the energy level splitting of the system), one obtains the quantum-optical master equation, which is of Gorini-Kossakowski-Sudarshan-Lindblad (GKSL) form \cite{GoriniKossakowski1976,GLindblad1976,RadiationConsideredReservoir1998,CWGardiner00,breuerpetruccione,wallsQuantumOptics2008}. 
This form brings many advantages: it is time local, induces a dynamics that is completely positive and trace preserving, and it can efficiently be unravelled using stochastic quantum trajectory simulations \cite{Dalibard92,Kmolmer1993,GardinerZoller1992:1,GardinerZoller1992:2,NGisin1992:1,HCarmicheal1999,ADaley2014}. 
However, the assumption of ultraweak system-bath coupling is often not fulfilled, especially for large quantum systems, where the level splitting tends to decrease with system size \cite{MunroGardiner1996,HWichterich2007,vrajitoareaUltrastrongLightmatterInteraction2022}. 
Beyond ultraweak coupling one still finds time-local master equations, like the Redfield equation, but these are mostly not of GKSL form anymore. 
However, often these equations can still be brought to pseudo-Lindblad form, with a dissipator that resembles that of a GKSL equation, except for the fact that some terms of the dissipator possess negative prefactors \cite{AShajiECGSudarshan2005,Piilo2009,deVaga2017,CGneiting2020,TBecker2021,Groszkowski2022}. 
While pseudo-Lindblad equations can, in principle, violate the positivity of the density operator,  when applied outside their regime of validity \cite{VRomero1989,ASuarezIOppenheim92,PPechukas94,EGevaERosenman2000,AMCastilloDRReichman15,Strunz20}, they still allow for an unravelling in terms of quantum-trajectories \cite{strunzOpenSystemDynamics1999,strunzConvolutionlessNonMarkovianMaster2004,PiiloManiscalco2008,beckerQuantumTrajectoriesTimeLocal2023} and they can be approximated by a GKSL equation obtained from truncating the negative non-GKSL terms \cite{TBecker2021}. 
Both approaches perform better, when the weights of the negative terms in the dissipator are smaller. In Ref.~\cite{TBecker2021} we have seen already that the pseudo-Lindblad form of the Redfield equation is not unique and that, in particular, it is possible to alter the weight of the dissipators with negative prefactor. In view of the above approximation schemes, here a choice that minimizes these negative-weight terms is optimal.

Taking this last consideration as a starting point, in this paper, we systematically investigate transformations that leave the generator of pseudo-Lindblad equations unchanged, while changing the absolute and relative weights of their positive and negative dissipation terms. 
We show that these transformations allow for  bringing the equation into an optimal form with the negative dissipation terms having minimal weight. 
We then apply them to the Redfield equation, which can be brought to pseudo-Lindblad form \cite{TBecker2021}, and demonstrate that the minimal weight of the negative dissipation term previously found in Ref.~\cite{TBecker2021}, indeed, corresponds to the smallest possible value. 
We also determine the pseudo-Lindblad form of the Redfield equation in energy representation and  
discuss the role of the negative dissipation terms in the limit of ultraweak coupling.
In order to illustrate the advantage of the optimization, we show for the concrete example of an open extended Hubbard model that the convergence of a pseudo-Lindblad quantum jump unravelling \cite{beckerQuantumTrajectoriesTimeLocal2023} indeed dramatically improves, when using the optimization.
Finally, we also consider the Hu-Paz-Zhang master equation, which exactly solves a damped harmonic oscillator \cite{HuPazZhang1992,KarrleinGrabert97,Paz94,GrabertWeiss84}, and show that it can be brought to the same form as the Redfield equation and, therefore, also to pseudo-Lindblad form.

\section{Symmetry transformations of pseudo-Lindblad dissipators}
\label{sec:symmetry_pseudo_Lind}
Our starting point is the pseudo-Lindblad equation for the density operator $\rho$ of the system (in units, where $\hbar=1$),
\begin{align}
	\partial_t \varrho(t) =-\i [H,\varrho(t)] +  \mathcal{D}(A_+)[\varrho(t)] - \mathcal{D}(A_-)[\varrho(t)] 
	\label{eq:pseudo_Lindblad}.
\end{align}
It is characterized by a Hamiltonian $H$, describing the coherent evolution, and two dissipation terms given by GKSL superoperators $\mathcal D(A)[\varrho(t)] = A\varrho(t) A^\dagger -\frac{1}{2} \{ A^\dagger A,\varrho(t)\}$, with prefactors $\pm1$ and respective jump operators
$A_+$ and $A_-$. The weights of the dissipation terms are absorbed into the jump operators and all operators might be time dependent. 
The motivation for considering the equation with one positive and one negative dissipation term is motivated by the fact that this is the form obtained both from the Redfield equation (see Ref.~\cite{TBecker2021} and Sec.~\ref{sec:pseudo_lind_redfield} below) and the Hu-Path-Zhang master equation (see Sec.~\ref{sec:pseudo_lindblad_dHO}). 
A generalization to the case of more than one positive and one negative jump operator is possible, but generally would require numerical optimization. A simpler scenario with multiple dissipators is a sum of pairs, $\sum_\ell \mathcal D(A_{\ell+}) - \mathcal D(A_{\ell-})$, to be optimized individually (cf.~\cref{sec:PLQT}).
Generally, pseudo-Lindblad equations are known to describe non-Markovian dynamics \cite{Piilo2009, deVaga2017}.
In the respective parameter regime where a pseudo-Lindblad equation is valid, it does not lead to (relevant) violations of positivity \cite{AShajiECGSudarshan2005}. 
This is seen both for phenomenological pseudo-Lindblad equations \cite{DCruscinskiJPiiloWStrunz2017} as well as for microscopic ones as discussed further below. 

It is known that (pseudo-)Lindblad equations are invariant under certain transformations that change the jump operators as well as the Hamiltonian \cite{breuerpetruccione,CollaBreuer2022}. 
On the one hand, the separation between the coherent part, as it is described by the Hamiltonian $H$, and the dissipative part, characterized by the jump operators, is not unique and can be chosen to minimize the dissipation term \cite{CollaBreuer2022}. In this paper, we focus on another possible transformation that modifies the dissipator alone. It is mentioned (but not exploited) in Ref.~\cite{CollaBreuer2022}. 
Here, we construct it explicitly and use it to address a different problem, namely the possibility of reducing the weight of the negative relaxation term. As mentioned above, minimizing this weight is optimal for quantum-trajectory simulations \cite{beckerQuantumTrajectoriesTimeLocal2023} and GKSL approximations based on truncating the negative term \cite{TBecker2021}.

As a first step, we write the dissipation term as a bilinear form
\begin{align}\label{eq:PLD}
	\mathcal{D}[\varrho(t)]
	&=\mathcal{D}(A_+)[\varrho(t)] - \mathcal{D}(A_-)[\varrho(t)] \notag \\&= \sum_{i,j=+,-} \sigma^z_{ij} \Bigg[ A_i \varrho(t) A_j^\dagger - \frac{1}{2} \Big\{ A_j^\dagger A_i, \varrho(t) \Big\}\Bigg] ,
\end{align}
with $\sigma^z_{ij}$ being the elements of the Pauli-z-matrix.
In the form above, where two operators $A_i$ and $A_j$ act in this particular form on the density matrix, the coefficient matrix is referred to as Kossakowski matrix \cite{GoriniKossakowski1976}. 
Here, the Kossakowski matrix is diagonal, but, in the next section, we generalize the concept to non-diagonal Kossakowski matrices. 

We are interested in symmetry transformations $W\in \mathbb C^{2,2}$ that leave the Kossakowski matrix invariant, 
\begin{align}
	\sigma^z = W \sigma^z W^\dagger .
	\label{eq:pseudo_unitary}
\end{align}
This defines a reference frame, with a different pair of positive and negative jump operators, i.e., 
\begin{align}\label{eq:Wtrans}
(A_+, A_-) \rightarrow (A_+, A_-) W \equiv (A^W_+, A^W_- ) ,
\end{align}
which is a product of the row vector $(A_+, A_-)$ and the matrix $W$. 
Here, the Kossakowski matrix is diagonal with the entries $1$ and $-1$, and, consequently, \cref{eq:pseudo_unitary} is the definition of a two dimensional pseudo-unitary matrix~$W$. 

For the components,
\begin{align}
	W = \left(\begin{array}{rr}
		W_{11} &  W_{12} \\
		W_{21} &  W_{22} \\
	\end{array}\right),
\end{align}
the symmetry condition $\sigma^z = W \sigma^z W^\dagger$ yields
\begin{align}
	1 & = |W_{11}|^2 - |W_{12}|^2 , \\
	1 & = |W_{22}|^2 - |W_{21}|^2 , \\
	0 &= W_{21}W_{11}^* - W_{22} W_{12}^*, \label{eq:third_constraint}
\end{align}
which fixes four out of eight real entries in $W$. 
Note that the latter identity gives two constraints, i.e., one condition from the real part and another condition from the imaginary part. 
The first two conditions are satisfied by the parametrization
\begin{align}
	W = \left(\begin{array}{rr}
		e^{\i \mathrm{arg}(W_{11})} \cosh(w_1) &  e^{\i \mathrm{arg}(W_{12})} \sinh(w_1) \\
		e^{\i \mathrm{arg}(W_{21})} \sinh(w_2) &  e^{\i \mathrm{arg}(W_{22})} \cosh(w_2) \\
	\end{array}\right) ,
\end{align}
where $\mathrm{arg}(W_{ij})$ denotes the complex phase for the entry $W_{ij} = |W_{ij}| e^{\i \mathrm{arg}(W_{ij})}$ and where $w_i\ge0$.
Further, from the third condition in \cref{eq:third_constraint}, by taking the absolute value, it follows that $|W_{21}|/|W_{22}| = |W_{12}|/|W_{11}|$, which is satisfied for $w = w_1 = w_2$.
As a final step, out of the four complex phases one complex phase is fixed, since \cref{eq:third_constraint}  gives
$ \mathrm{arg}(W_{11})+ \mathrm{arg}(W_{22})  - \mathrm{arg}(W_{12})-\mathrm{arg}(W_{21})=2\pi n$ with integer $n$. All in all, we obtain the the most general parametrization of the symmetry transformation $W$
\begin{align}
	W = e^{\mathrm i \alpha} \left(\begin{array}{rr}
		e^{\i \frac{\varphi + \beta}{2}} \cosh(w) &  e^{\i \frac{\varphi - \beta}{2}} \sinh(w) \\
		e^{-\i \frac{\varphi - \beta}{2}}\sinh(w) &  e^{-\i \frac{\varphi + \beta}{2}} \cosh(w) \\
	\end{array}\right) ,
	\label{eq:symmetry_matrix}
\end{align}
with four real parameters given by the argument $w\ge0$ of the hyperbolic functions and three complex phases $\varphi, \alpha$, and $\beta$.

By carrying out the matrix product (\ref{eq:Wtrans})
we obtain a class of equivalent pseudo-Lindblad representations, 
\begin{align}
\mathcal{D}[\rho]=\mathcal{D}(A^W_+) )[\varrho(t)] - \mathcal{D}(A^W_-)[\varrho(t)],
\end{align} 
with jump operators
\begin{align}
	A^W_+ &=e^{\i \alpha + \i \frac{\beta}{2}} \left( e^{\i \frac{\varphi}{2}}\cosh(w) A_+ + e^{-\i \frac{\varphi}{2}} \sinh(w) A_- \right) , \label{eq:pos_jump_pseudo}\\
	A^W_- &=e^{\i \alpha - \i \frac{\beta}{2}} \left( e^{\i \frac{\varphi}{2}}\sinh(w) A_+ + e^{-\i \frac{\varphi}{2}} \cosh(w) A_- \right)  , \label{eq:neg_jump_pseudo}
\end{align}
for the positive and negative relaxation weight, respectively.
As a consequence of \cref{eq:symmetry_matrix}, and as can be verified easily, one has $A^W_+ A^{W\dagger}_+ - A^W_- A^{W\dagger}_- = A_+ A^{\dagger}_+ - A_- A^{\dagger}_-$, which is invariant under the transformation. 
It is evident that the global phases of the jump operators drop out of the dissipators, so that we can choose the phases $\alpha$ and $\beta$ to be zero. Hence, they can be omitted, such that there are only two relevant symmetry parameters $w$ and $\varphi$. 
\begin{align}
	A_+^{(w,\varphi)}&= \left( e^{\i \frac{\varphi}{2}}\cosh(w) A_+ + e^{-\i \frac{\varphi}{2}} \sinh(w) A_- \right) , \label{eq:pos_jump_wphi}\\
	A_-^{(w,\varphi)} &= \left( e^{\i \frac{\varphi}{2}}\sinh(w) A_+ + e^{-\i \frac{\varphi}{2}} \cosh(w) A_- \right)  , \label{eq:neg_jump_wphi}
\end{align}
Using the freedom to choose the symmetry parameters, we can minimize the weight of the negative jump operator. In the following, we will do that explicitly both for the Redfield equation and the Hu-Paz-Zhang master equation. 

\section{Application to the Redfield equation }
\label{sec:pseudo_lind_redfield}
In this section we apply the symmetry transformation to the Redfield equation. 
Based on the total system-bath Hamiltonian $H_\mathrm{tot} = H_\mathrm{S} + H_\mathrm{SB} + H_\mathrm{B}$, for weak system-bath coupling $H_\mathrm{SB} = S \otimes (B^\dagger + B)$, one obtains the Redfield equation for the reduced dynamics of the system. 
It can be written in the form
\begin{align}
		\partial_t \varrho(t) =- \i [H_\mathrm{S} + H_\mathrm{LS}, \varrho(t)] + \mathcal D^\mathrm{Red}[ \varrho(t)] , 
\end{align}
with Lamb-shift Hamiltonian $H_\mathrm{LS}$ [\cref{sec:Redfield_eq}] and Redfield dissipator
\begin{align}
	\mathcal D^\mathrm{Red}[\varrho(t) ] = \sum_{i,j=1}^2 \sigma^x_{ij} \Bigg[ S_i \varrho(t) S_j^\dagger - \frac{1}{2} \Big\{ S_j^\dagger S_i, \varrho(t) \Big\}\Bigg] ,
\end{align}
with non-diagonal Kossakowski matrix $\sigma^x$ and operators $S_1=S$ and $S_2 = \mathbb{S}$. Here  $\mathbb{S}$ denotes the the convoluted coupling operator:
\begin{align}
	\mathbb S = \int_0^\infty C(t)  S(-t) dt,
	\label{eq:convolution_operator}
\end{align}
where the bath correlation function $C(t) = \expval{  B^\dagger(t) B(0) + B(t) B^\dagger(0)}$ defined as the expectation values of the bath operators in the interaction picture  see [\cref{sec:Redfield_eq}]. 

By diagonalizing the Kossakowski matrix, one obtains a representation of the Redfield equation with new jump operators that act symmetrically on the state. 
We consider the unitary transformation,
\begin{align}
	U = \frac{1}{\sqrt 2}  \left(\begin{array}{rr}
	1 &  -1\\
	1 & 1 \\
\end{array}\right) ,
\end{align}
which defines a rotated frame in which the Kossakowski matrix becomes diagonal
\begin{align}
\sigma^x \to U^\dagger \sigma^x U = \sigma^z,
 \end{align}
with the new set of jump operators
\begin{align}
	(S, \mathbb S) \to  (S, \mathbb S)\, U \equiv (A_+, A_-),
\end{align}
corresponding to the relaxation weights $1$ and $-1$ for $A_+$ and $A_-$, respectively. 

Here, it is immediately clear that the Kossakowski matrix has one negative eigenvalue and therefore the Redfield equation is not of GKSL form, but instead of pseudo-Lindblad form, 
\begin{align}
	\mathcal{D}^\mathrm{Red} = \mathcal{D}(A_+) - \mathcal{D}(A_-).
	\label{eq:RedD}
\end{align}
This equation is of the form of Eq.~(\ref{eq:PLD}), which was the starting point of the previous section.  It was obtained without invoking a basis representation. Below, we will explicitly consider the basis of energy eigenstates and analytically solve the eigenvalue problem for the resulting Kossakowski matrix.  

The diagonal representation of the Redfield dissipator in pseudo-Lindblad form (\ref{eq:RedD}) allows to apply the symmetry transformation of the previous section. 
Altogether, the transformation is given by a product of the diagonalization $U$ and the symmetry transformation $W$, $(S,\mathbb S) \to (A_+^W,A_-^W) = (S,\mathbb S)\, U W$. 
By carrying out the matrix product the new jump operators are
\begin{align}
	A_+^{(w,\varphi)} = \frac{1}{\sqrt{2}} \Big( &[e^{\i \frac{\varphi}{2}} \cosh(w) - e^{- \i \frac{\varphi}{2}} \sinh(w)] S \\ \notag
	+ &[e^{\i \frac{\varphi}{2}} \cosh(w) + e^{- \i \frac{\varphi}{2}} \sinh(w)] \mathbb S \Big) , \\ 
	A_-^{(w,\varphi)} = \frac{1}{\sqrt{2}} \Big( -&[e^{\i \frac{\varphi}{2}} \cosh(w) - e^{- \i \frac{\varphi}{2}} \sinh(w)]^* S \notag \\ 
	+&[e^{\i \frac{\varphi}{2}} \cosh(w) + e^{- \i \frac{\varphi}{2}} \sinh(w)]^* \mathbb S \Big) .
\end{align}
Using a different parametrisation, i.e., $\cosh(w) - \sinh(w) = \lambda$ and $\cosh(w) + \sinh(w) = 1/\lambda$, one arrives at the equivalent representation
\begin{align}
	A^{(\lambda,\varphi)} _\sigma =A^{(w,\varphi)} _\sigma = \frac{1}{\sqrt{2\cos \varphi}}  \Big[ \sigma \lambda e^{\i \sigma  \frac{\varphi}{2}}\ \hat{S} + \frac{1}{\lambda}  e^{- \i \sigma \frac{\varphi}{2}}\ \hat{\mathbb{S}} \Big] , 
	\label{eq:jumpoperators}
\end{align}
with $\sigma=+1,-1$. 
This form was obtained previously in Ref.~\cite{TBecker2021} and is particularly useful for minimizing the weight associated with the jump operator $A_-$. 
Here, we reproduce this result, however, by considering the most general symmetry transformation and, thus, demonstrate that this is indeed the most general form. 

The jump operators $A^{(\lambda,\varphi)}_\sigma$ can, equivalently be obtained by first applying a symmetry transformation and then diagonalizing Kossakowski matrix. 
This is shown explicitly in \cref{sec:symmetry_Redfield}, where we start from transformations $\Lambda \in \mathbb C^{2,2}$ that leave the Kossakowski matrix $\sigma^x$ invariant,
\begin{align}
	\sigma^x = \Lambda \sigma^x \Lambda^\dagger . 
\end{align}
The new jump operators in \cref{eq:jumpoperators} are then given by $(A^{(\lambda,\varphi)}_+, A^{(\lambda,\varphi)}_-) = (S,\mathbb S)\, \Lambda U$. 

\section{Minimize the negative weight}
\label{sec:minimize_neg_weight}
In this section we minimize the weight of the jump operator $A_-^{(\lambda,\varphi)}$ for the Redfield equation by varying the parameters $\lambda$ and $\varphi$. The results of this section review previous work described in Ref.~\cite{TBecker2021}. Now, however, we know that  we indeed find the minimal possible weight, since Eq.~(\ref{eq:jumpoperators}) describes the most general variation of the jump operators. 

To begin with, we assign weights to the jump operators $A^{(\lambda,\varphi)}_\sigma$ in \cref{eq:jumpoperators}, which we define as the Frobenius norm 
\begin{align}
\lVert A_\sigma^{(\lambda,\varphi)}\rVert^2 = \tr\left\{A_\sigma^{(\lambda,\varphi)} (A_\sigma^{(\lambda,\varphi)})^\dagger\right\}.
\end{align}
The choice of the norm is not unique, though we expect qualitatively similar results for other norms. Here we have picked the Frobenius norm, which is obtained from the scalar product $(\varrho_1, \varrho_2 )\mapsto \tr(\varrho_1 \varrho_2^\dagger)$, because of the favorable property that it fulfills the additivity relation $\lVert A + B \rVert^2 = \lVert A \rVert^2 + 2\, \mathrm{Re} \tr (AB^\dagger) + \lVert B \rVert^2 $, which allows us to minimize it for the jump operators analytically. 
The weights of the jump operators are then given by
\begin{align}
	\lVert A^{(\lambda,\varphi)}_\sigma \rVert^2 = &\sigma \mathrm{Re}\tr(\mathbb SS)  - \tan(\varphi)\mathrm{Im}\tr(\mathbb SS)   \notag \\ 
	&+ \frac{1}{2\cos(\varphi)} \left[ \lambda^2 \lVert S\rVert^2+ \frac{1}{\lambda^2} \lVert \mathbb S\rVert^2 \right]  ,
	\label{eq:relaxrates}
\end{align}
and change under the symmetry transformation $W$. 
We define the optimal transformation by the symmetry parameters $\lambda_\mathrm{min}$ and $\varphi_\mathrm{min}$ that minimize the negative weight $\lVert A_\sigma^\mathrm{min}\rVert$. 

It is interesting to observe that the difference of  the positive and the negative weight, 
\begin{align}  
	\lVert A^{(\lambda,\varphi)}_+ \rVert^2 -  \lVert A^{(\lambda,\varphi)}_- \rVert^2 = 2 \mathrm{Re} \tr(\mathbb S S),
	\label{eq:net_weight}
\end{align}
are independent of $\lambda$ and $\varphi$ and, thus,  invariant under the symmetry transformation.
We will see below that it is this weight difference, which determines the rates in the quantum-optical master equation, which is derived from the Redfield equation. As a result, the quantum-optical master equation is also invariant under the transformation (\ref{eq:jumpoperators}), as expected. Furthermore, the fact that the rate difference (\ref{eq:net_weight}) is fixed, implies that by minimizing the absolute weight $\lVert A_-^{(\lambda,\varphi)}\rVert^2 $ with respect to the symmetry transformations $W$, at the same time we minimize also the relative weight of the negative dissipation,  $\lVert A_-^{(\lambda,\varphi)}\rVert^2 /\lVert A_+^{(\lambda,\varphi)}\rVert^2 $, as well as the total dissipation weight $\lVert A_+^{(\lambda,\varphi)}\rVert^2 + \lVert A_-^{(\lambda,\varphi)}\rVert^2 $.

Interpreting $\lambda^2$ and $\varphi$ as variational parameters we find the minimal weights with respect to the the Frobenius norm \cite{TBecker2021}
\begin{align}
	\lVert A^\mathrm{min}_\sigma \rVert^2 = \sigma \mathrm{Re}\tr(\mathbb SS) + \sqrt{\lVert S\rVert^2 \lVert \mathbb S\rVert^2 - [\mathrm{Im}\tr (\mathbb SS)]^2}
	\label{eq:optimal_weights}
\end{align}
for the optimal symmetry parameters
\begin{align}  
	\lambda_\mathrm{min}^2 & = \frac{ \lVert \mathbb{S} \rVert }{  \lVert  S\rVert}, \label{eq:optimal_lambda} \\
	\sin\varphi_\mathrm{min} & = \frac{\mathrm{Im} \tr(\mathbb S S)}{  \lVert \mathbb{S} \rVert  \lVert  S\rVert} . \label{eq:optimal_phi}
\end{align}

Let us now evaluate the jump operators and their weights in the energy-eigenbasis of the Hamiltonian,  $H|k\rangle=E_k|k\rangle$.
For this purpose, we need to evaluate the traces $ \lVert  S\rVert$, $\tr(\mathbb S S)$, and $ \lVert \mathbb{S} \rVert$. One finds matrix elements $\matrixel{q}{S}{k} = S_{qk}$ and $\matrixel{q}{\mathbb S}{k} = G(\Delta_{qk}) S_{qk}$, where $G(\Delta_{qk})$ is the coupling density evaluated at the energy level splitting $\Delta_{qk} = E_q - E_k$ see [\cref{sec:temp_dep_coupling_dens}].
For the traces, one obtains $\lVert S\rVert^2 = \sum_{qk} |S_{qk}|^2$, $\tr(\mathbb SS) = \sum_{qk} G(\Delta_{qk})$ and $\lVert \mathbb S\rVert^2 = \sum_{qk} |S_{qk}|^2 |G(\Delta_{qk})|^2$. 
Here, it is handy to introduce the average 
\begin{align}
	\expval{x}_S \equiv \frac{1}{\lVert S\rVert^2} \sum_{qk} |S_{nm}|^2 x(\Delta_{nm}), 
	\label{eq:average}
\end{align}
such that $\tr(\mathbb SS) = \lVert S\rVert^2 \expval{G}_S$ and $\lVert \mathbb S\rVert^2 = \lVert S\rVert^2 \expval{|G|^2}_S$.
The optimal parameters are $\lambda^2_\mathrm{min} = \sqrt{\expval{|G|^2}_S}$ and  $\sin \varphi_\mathrm{min} = \expval{G''}_S/ \sqrt{\expval{|G|^2}_S}$ and the minimal weights can be written as 
\begin{align}
	\lVert A^\mathrm{min}_\sigma \rVert^2 = \lVert S\rVert^2 \expval{G'}_S \Big(\sigma  + \sqrt{1 + \frac{V[G']_S+ V[G'']_S}{\expval{G'}_S^2}  }\Big) ,
	\label{eq:optimal_weights_basis}
\end{align}
where we introduce the variance $V[x]_S \equiv \expval{x^2}_S - \expval{x}_S^2$.

In summary, we find model independent expressions for the optimal weights, Eq.~(\ref{eq:optimal_weights}), and the corresponding transformation parameters, \cref{eq:optimal_lambda,eq:optimal_phi}, and we can express them also in the energy basis. These are found to minimize the absolute and relative weight of $A_-$ as well as of the total dissipation with respect to the symmetry transformations $W$.
 
\section{Diagonalization of the Redfield dissipator in energy representation}
In this section (as well as in Appendix~\ref{sec:analytical_diagonalization}), we derive a pseudo-Lindblad form for the Redfield equation using a completely different approach. Namely, we analytically diagonalize the Kossakowski matrix of the Redfield equation in energy representation using a unitary transformation. 

Usually, the Redfield equation is represented in the energy eigenbasis of the system, with eigenstate $\ket{q}$ and corresponding energy $E_q$, with $q=1,\dots,D$.  
The eigenbasis representation is relevant for the numerical implementation of the Redfield equation. 
Furthermore, it is the basis in which the rotating-wave approximation is introduced to derive the quantum-optical master equation \cite{CWGardiner00,breuerpetruccione,weissQuantumDissipativeSystems2012}. 
In the energy basis, the Redfield dissipator takes the form [see \cref{sec:analytical_diagonalization}]
\begin{align}
		\mathcal{D}^\mathrm{Red}[\varrho(t)]  =  \sum\limits_{qk, q'k'} &S_{qk} S_{k'q'} M_{(qk),(q'k')} \notag \\& \Big[ \hat{L}_{qk} \varrho(t) \hat{L}_{q'k'}^\dagger-  \frac{1}{2} \Big\{ \hat{L}_{q'k'}^\dagger \hat{L}_{qk} ,\varrho(t)  \Big\} \Big],
	\label{eq:redfield_in_components_main}
\end{align}
with the Hermitian Kossakowski matrix $M$, 
\begin{align}
	M_{(qk),(q'k')} = G(\Delta_{qk}) + G(\Delta_{q'k'})^* ,
\end{align}
where $G(\Delta)$ is again the complex coupling density. 
As a novel result, we analytically solve the eigenvalue problem $M v^\sigma = g_\sigma v^\sigma$ of the Kossakowski matrix in Appendix~\ref{sec:analytical_diagonalization}. In this way, we find a diagonal representation of the Redfield dissipator
\begin{align}
\mathcal{D}^\mathrm{Red} = \mathcal{D}(A_+^0) - \mathcal{D}(A_-^0),
\end{align}
with jump operators
\begin{align}
	A_\sigma^0
	&= \sum\limits_{qk} \sqrt{|g_\sigma|} \frac{v^\sigma_{qk}}{\lvert v^\sigma \rvert} S_{qk} \dyad{q}{k} , \notag \\
	&= \frac{1}{\sqrt{2\cos \varphi_0}}  \left( \frac{1}{\lambda_0} e^{\mathrm i \sigma \frac{\varphi_0}{2}} \mathbb S + \sigma \lambda e^{-\mathrm i \sigma \frac{\varphi_0}{2}}  S \right).
	\label{eq:jump_operators_eigenbasis}
\end{align}
Qualitatively, the jump operators can be written in the general form of \cref{eq:jumpoperators}, however, with parameters
\begin{align}
	\lambda^2_0 & = \sqrt{\expval{|G|^2}}  , \label{eq:eigenbasis_lambda}\\
	\sin \varphi_0 & = \frac{\expval{G''}}{\sqrt{\expval{|G|^2}}  } \label{eq:eigenbasis_phi} .
\end{align}
In difference to the optimal parameters in \cref{eq:eigenbasis_lambda,eq:eigenbasis_phi}, here, the average is defined as $\expval{x} \equiv \frac{1}{D^2} \sum_{qk} x(\Delta_{qk})$, which is independent of the coupling operator $S$.
Therefore the parameters are not optimized for the coupling operator $S$. 
The weights associated with these jump operators are of the same form as in \cref{eq:optimal_weights_basis}, but with the optimal average $\expval{\cdot}_S$ replaced by the bare average $\expval{\cdot }$ [cf.~\cref{sec:analytical_diagonalization}].

\section{Impact of optimization on convergence of quantum-trajectory unravelling}
\label{sec:PLQT}
\begin{figure}[b]
	\includegraphics{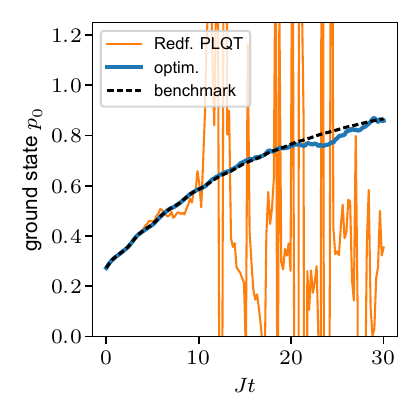}
	\caption{Dynamics of the ground-state population predicted by $N=10^4$ pseudo-Lindblad quantum trajectories for the Redfield equation without optimization in thin orange and with optimization thick blue. Direct integration of the Redfield equation as benchmark in dashed black. We consider an extended Hubbard model of $M=4$ sites filled with two fermions, tunneling strength $J$ and nearest-neighbor interaction $V/J=1$. All sites are coupled ($\gamma=0.25$) to a thermal bath of temperature $T/J=0.1$ with Drude cutoff frequency $\omega_D/J=1$.}
	\label{fig:figure1}
\end{figure}

\begin{figure}[t]
	\includegraphics{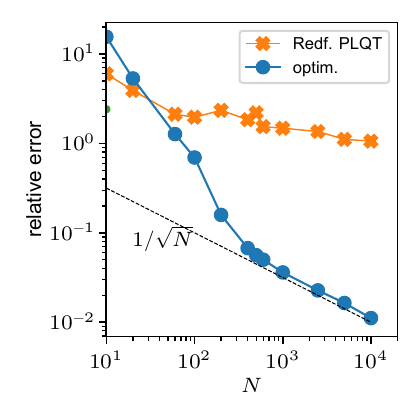}
	\caption{Time averaged relative error $| p_0^\text{sample} - p_0 |/p_0$ of the ground-state population predicted by the PLQT algorithm as a function of $N$ in double logarithmic scale. 
	We compare the convergence of PLQT without optimization (orange crosses) to that with optimization (blue bullets). For each data point we average the relative error over sufficiently many independent 	
	realizations of $N$-trajectory ensembles, so that sample-specific fluctuations of the relative error become small.}
	\label{fig:figure2}
\end{figure}

Quantum master equations represented in pseudo-Lindblad form, such as for the Redfield equation in \cref{eq:RedD}, can be unravelled by means of pseudo-Lindblad quantum trajectories (PLQT) \cite{beckerQuantumTrajectoriesTimeLocal2023}. 
This can be of advantage compared to the direct integration of the master equation, since a quantum trajectory unravelling allows for the simulation of larger systems because the memory required for simulating a state vector growths linear with the Hilbert space dimension and not quadratically as for the density matrix. 

In the PLQT approach, the open system is described by an ensemble of $N$ pure states $|\psi_n(t)\rangle$, whose time evolution is given by the dynamics generated by an effective non-hermitian Hamiltonian, which is interrupted at random times by randomly chosen quantum jumps described by the jump operators of the peseudo-Lindblad equation. 
Different from the standard quantum-jump unravelling of Lindblad equations with only positive weights in the dissipator, in the PLQT approach each state $|\psi_n(t)\rangle$ is accompanied by a classical sign bit $s_n(t)$, which at each time $t$ either takes the value $+1$ or $-1$. 
The sign bits are initialized so that $s_n(0)=+1$ for all trajectories $n$ and they flip sign, whenever a quantum jump with negative weight  (i.e.\ of $A_-$-type) occurs. 
At the end, the density matrix of the system is approximated by the sample average \cite{beckerQuantumTrajectoriesTimeLocal2023}
\begin{equation}
\rho(t)\simeq \frac{1}{\mathcal{N}(t)}\sum_{n=1}^N s_n(t)|\psi_n(t)\rangle\langle \psi_n(t)|\equiv \rho_\text{sample}(t),
\end{equation}
with the normalization factor defined by $\mathcal{N}(t)=\sum_{n=1}^N s_n(t)\langle \psi_n(t)|\psi_n(t)\rangle$. This unravelling becomes accurate for sufficiently large $N$ and approaches the exact density matrix in the limit $N\to\infty$. 

The convergence of the unravelling with respect to the number of trajectories $N$ depends, however, sensitively on the fraction of sign bits $s_n(t)$ having negative values. This can be illustrated by looking at a specific observable, namely the probability $p_\alpha(t)$ for being in a particular (but arbitrary) state $|\alpha\rangle$, 
\begin{align}
	p_\alpha(t) &=\tr\{\rho(t) |\alpha\rangle\langle\alpha|\}  \notag \\
	&\simeq \frac{1}{\mathcal{N}(t)}\sum_{n=1}^N s_n(t)|\langle\alpha|\psi_n(t)\rangle|^2.
	 \notag \\
	&= \frac{1}{N}\sum_{n=1}^N p_\alpha^{(n)}(t) \equiv p_\alpha^\text{sample}(t),
\end{align}
with
\begin{align}
	p_\alpha^{(n)}(t) = s_n(t)|\langle\alpha|\psi_n(t)\rangle|^2 \frac{N}{\mathcal{N}(t)}.
\end{align}
Let $f_\alpha(p;t)$ be the probability distribution at time $t$ from which the $N$ values $p_\alpha^{(n)}(t)$ of our sample are drawn \footnote{Note that in actual numerical calculations based on the PLQT algorithm, this distribution is never constructed explicitly. The elements of the sample, $p_\alpha^{(n)}(t)$ are obtained from the quantum trajectories given by $|\psi_n(t)\rangle$ and $s_n(t)$, which are obtained by integrating a stochastic equation of motion, as described above.
Nevertheless, we can define the distribution $f_\alpha(p;t)$ from which the elements $p_\alpha^{(n)}(t)$ are drawn and, in principle, we could also reconstruct it numerically from the quantum jump approach}.
The PLQT is constructed, so that its expectation value is given by the exact solution, $\mu_\alpha(t) = \int p f_\alpha(p;t) dp = p_\alpha$.
The variance $\sigma_\alpha^2(t)= \int (p-p_\alpha)^2f_\alpha(p;t) dp$, in turn, varies with time and depends (as we will discuss below) on the strength of the negative-weight jump operators~$A_-$. It directly influences the expected deviation $\delta p_\alpha^\text{sample}$ of the approximate result $p_\alpha^\text{sample}(t)$ from the exact value $p_\alpha(t)$.
Namely, for a randomly drawn finite sample of $N$ elements, the standard deviation of the sample average, $p_\alpha^\text{sample}(t)$, from its expectation value, $p_\alpha(t)$, is given by $\delta p_\alpha^\text{sample} = \sigma_\alpha(t)/\sqrt{N}$. 

For an actual Lindblad equation, having only positive-weight jump operators, we find $p_\alpha^{(n)}(t)\ge0$ for all trajectories.
Thus, the probability distribution is non-zero for positive $p$ only, i.e., $f_\alpha(p;t)=0$ for $p<0$.
As a consequence, the standard deviation $\sigma_\alpha$ of this distribution is typically not larger than its expectation value, unless the distribution possesses a large tail.
Thus the expected relative sampling error $\delta p_\alpha^\text{sample}/p_\alpha$ is typically of the order of $N^{-1/2}$. 
For a pseudo-Lindblad equation, having also negative-weight quantum jumps, this is approximately true for early times $t\ll\tau$, when the majority of the sign bits have not yet flipped.
Here $\tau$ shall be the average time on which a negative-weight ($A_-$-type) quantum jump occurs.
As a result, most of the values $p_\alpha^{(n)}(t)$ are still positive, which implies that the distribution $f_\alpha(p;t)$ is small for negative $p$.
At times $t\gtrsim\tau$, however, positive and negative signs $s_n(t)$ eventually become equally likely.
This implies that the distribution $f_\alpha(p;t)$ acquires non-negligible contributions for $p<0$, allowing for $\sigma_\alpha(t)>p_\alpha$ even without large tails in the distribution, so that for the same relative accuracy $\delta p_\alpha^\text{sample}/p_\alpha$ much larger sample sizes $N$ are needed.  
Note that the time scale $\tau$ for sign flips is \emph{not} a property of the physical system, but is rather a property of the PLQT algorithm.
It depends directly on the strength of the negative-weight jump operators $A_-$.
Thus, by minimizing this weight, we can optimize the convergence of the unravelling with respect to $N$. 

For demonstration, we consider an extended Hubbard model for two spinless fermions on a one dimensional chain of $M=4$ sites described by the Hamiltonian
\begin{align}
	H_\mathrm{S} = -J \sum_{\ell =1}^M \big( a_\ell^\dagger a_{\ell+1} + a_{\ell+1}^\dagger a_\ell \big) + V \sum_{\ell=1}^{M}  a_\ell^\dagger a_{\ell}a_{\ell+1}^\dagger a_{\ell+1},
\end{align}
with tunneling strength $J$ and nearest-neighbor interaction ${V/J=1}$.
The system bath Hamiltonian shall be given by 
\begin{align} 
	H_\mathrm{SB} = \sqrt{\gamma} \sum_{\ell =1}^{M}  a^\dagger_\ell a_\ell \otimes (B^\dagger_\ell + B_\ell),
\end{align}
with equal coupling strength $\gamma$ to identical thermal baths of temperature $T$.
For a pure Ohmic spectral density $J(\Delta) = \Delta/(1+\Delta^2/\omega_D^2)$ with cutoff frequency $\omega_D$, the convoluted coupling operators are expressed in the system's eigenbasis 
\begin{align} 
	\mathbb S_\ell  = \sum_{qk} \frac{J(\Delta_{qk})}{\exp(\Delta_{qk}/T) - 1} \matrixel{q}{ S_\ell}{k} \ketbra{q}{k} ,
\end{align} 
with energy level splittings $\Delta_{qk} = E_q - E_k$. 

In \cref{fig:figure1}, we obtain the dynamics of the ground-state population $p_0^\text{sample}(t)$ for $N=10^4$ quantum trajectories, both for the optimized pseudo-Lindblad equation (thick blue line) as well as without optimization, for $\lambda=1$ and $\varphi=0$ in \cref{eq:jumpoperators}, (thin orange line). 
We compare the results to a benchmark given by the exact solution of the full master (black dashed line).  
We can see that for the non-optimized case, the convergence breaks down roughly at times $J \tau =10$. 
In turn, the unravelling provides an accurate description within the whole time interval plotted (i.e.~at least up to $J \tau = 30$). 
In \cref{fig:figure2}, we compare the relative error (averaged over the evolution time) for both the optimized and the non-optimized equation as a function of $N$. 
We can see that while the former (blue bullets) becomes small and approaches the expected $1/\sqrt{N}$ behaviour for $N \sim 300$, the latter (orange crosses) remains large even for $N=10^4$. 
Thus, we can clearly see that the optimization procedure described above is of immediate relevance for the efficient simulation of open quantum systems described by pseudo-Lindblad equations like the Redfield equation.

\section{Ultraweak coupling limit and the relation to detailed balance}
\label{sec:detailed_balance}

For ultraweak system-bath coupling, the Redfield equation can be reduced to the quantum-optical master equation \cite{HCarmicheal1999,CWGardiner00,breuerpetruccione},
\begin{align}
	\begin{split}
		R^\mathrm{RWA}[\varrho(t)] =  &-\i [ H_\mathrm{LS}^\mathrm{RWA} , \varrho(t) ]  + \sum\limits_{qk} R_{kq}  \mathcal D(L_{qk}) [\varrho(t)]  ,
	\end{split}
	\label{eq:RWA}
\end{align}
by performing the rotating-wave approximation. 
Here, $L_{qk} =\dyad{q}{k}$ and $R_{kq} =  2\, G'(\Delta_{qk})|S_{qk}|^2 $ are the jump operators and the associated weights, respectively. 
Moreover, the Lamb-shift Hamiltonian reads $H_\mathrm{LS}^\mathrm{RWA} = \sum_{qk} G''(\Delta_{qk})|S_{qk}|^2 \dyad{k}$. 
Since the weights are positive, the quantum-optical master equation is of GKSL form and the the quantity $R_{kq} =  2\, G'(\Delta_{qk})|S_{qk}|^2 $ can be understood as the rate for quantum jumps from state $\ket{q}$ to~$\ket{k}$. 

For ultraweak coupling, the dynamics for the off-diagonal elements of the density matrix (coherences) decouples from the diagonal elements (populations). 
The coherences decay exponentially and the populations $p_q(t)=\matrixel{q}{\varrho(t)}{q}$ obey a Pauli-type classical rate equation, 
\begin{align}
	\partial_t p_q(t) = \sum_k [R_{qk} p_k(t) - R_{qk} p_q(t)]. 
	\label{eq:pauli_rate}
\end{align}
In equilibrium $\partial_t p_q(t)=0$, \cref{eq:pauli_rate} simplifies by noting $R_{kq}/R_{qk} = e^{-\beta \Delta_{qk}}$, which is known as the detailed balance condition, for which all terms vanish separately. 
Then the steady state is the canonical Gibbs state, $p^{(0)}_q = e^{-\beta E_q}/\sum_k( e^{-\beta E_k})$, which is known to be the exact solution in the limit of ultraweak coupling \cite{campisiFluctuationTheoremArbitrary2009,TalknerHaenggi2020}. 

Let us now show that both the negative and positive dissipation term in the pseudo-Lindblad representation of the Redfield dissipator  are crucial to obtain the detailed balance condition and with that the correct thermal steady state. 
Starting from the steady-state solution of the Redfield equation
\begin{align}
	0 = -\i [H_\mathrm{S} + H_\mathrm{LS}, \varrho] + \mathcal D^\mathrm{Red}[\varrho], 
	\label{eq:redfield_steady_state}
\end{align}
we make a series expansion for the steady state up to second order in the system-bath coupling, $\varrho \simeq \varrho^{(0)} + \varrho^{(2)}$. 
Evaluating \cref{eq:redfield_steady_state} for each order separately, one obtains the hierarchy of coupled equations,
\begin{align}
	0 &= -\i [H_\mathrm{S}, \varrho^{(0)}], \label{eq:steadystate_zeroth} \\
	0 &= -\i [H_\mathrm{S},\varrho^{(2)}] -\i [H_\mathrm{LS}, \varrho^{(0)}] + \mathcal D^\mathrm{Red}[\varrho^{(0)}] , \label{eq:steadystate_second}
\end{align}
where we have used that both the Lamb-shift and the dissipator are of second order with respect to the coupling.
From \cref{eq:steadystate_zeroth}, it follows that in zeroth order, the equilibrium state is diagonal in the energy basis, i.e., $\matrixel{q}{\varrho^{(0)}}{k} = \delta_{q,k} p_q^{(0)}$. 
The populations follow from \cref{eq:steadystate_second} in second order. 
We project \cref{eq:steadystate_second} to the populations and use that the commutator with a diagonal matrix is zero on the diagonal, such that $\matrixel{q}{[H_\mathrm{LS},\varrho^{(2)}]}{q} = \matrixel{q}{[H_\mathrm{S},\varrho^{(0)}]}{q} =0$ and find
\begin{align}
	0 &= \sum_k \matrixel{q}{\mathcal D^\mathrm{Red}[\dyad{k}{k}]}{q} \, p_k^{(0)} . 
\end{align}	

We now represent the Redfield dissipator in pseudo-Lindblad form, $\mathcal D^\mathrm{Red} = \mathcal D(A^{(\lambda,\varphi)}_+) - \mathcal D(A^{(\lambda,\varphi)}_-)$. 
Analogous to the quantum-optical master equation we find a rate equation 
\begin{align}
	0 = \sum_k \Big( &( |A^{(\lambda,\varphi)}_{+,{qk}}|^2 -  |A^{(\lambda,\varphi)}_{-, qk}|^2) p_k^{(0)} \notag \\  -& ( |A^{(\lambda,\varphi)}_{+, kq}|^2 -  |A^{(\lambda,\varphi)}_{-, kq}|^2) p_q^{(0)} \Big) , 
\end{align}
with rates given by the absolute squared matrix elements of the jump operators $ |A^W_{\sigma, qk}|^2 = |\matrixel{q}{A^W_\sigma}{k}|^2$. 
We see that the difference of the weights enters, which is independent of the symmetry transformation $W$. Thus, we can unambiguously identify the rates as
\begin{align}
R_{qk} = |A^{(\lambda,\varphi)}_{+,{qk}}|^2 -  |A^{(\lambda,\varphi)}_{-, qk}|^2, 
\end{align}
which can be reduced to the well-known Golden-rule type expression, giving rise to the detailed-balance condition. 

Another approximation of the Redfield equation giving rise to a GKSL master equation was discussed in Ref.~\cite{TBecker2021}. 
It is based on truncating the negative contribution of the Redfield equation in pseudo-Lindblad form. 
We find that this is valid beyond the ultraweak coupling limit, where especially for large bath temperatures and systems driven out of equilibrium it performs better than the quantum-optical master equation obtained from the rotating-wave approximation.  
However, as is also shown in Ref.~\cite{leePerturbativeSteadyStates2022}, the truncation of the negative rates eventually breaks the detailed balance condition and, consequently, the zeroth-order result differs from the canonical Gibbs state and, thus, does not provide the correct result in the limit of ultraweak coupling. 
Furthermore, the equilibrium populations become a complicated function of the symmetry parameters $\lambda_\mathrm{min}$ and $\varphi_\mathrm{min}$, the coupling operator $S$, as well as the complex coupling density $G(\Delta)$.
All this might be seen as a drawback of the truncation approximation and has to be considered for consistent results in the ultraweak coupling regime. 
However, it also shows the relevance of the negative contribution.
In particular, here, for ultraweak coupling it is shown that negative relaxation weights do not violate the positivity of the population, but on the contrary are essential for a consistent weak coupling limit. 
Thus, developing efficient methods to treat pseudo-Lindblad equations, including the negative terms are crucial. 
One example is the the pseudo-Lindblad quantum-trajectory approach \cite{beckerQuantumTrajectoriesTimeLocal2023}, which is found to become more efficient for smaller  weights of the negative dissipation term.

\section{Pseudo-Lindblad representation of the exact Hu-Paz-Zhang master equation}
\label{sec:pseudo_lindblad_dHO}
The damped harmonic oscillator with mass $M$ and frequency $\Omega$ is a paradigmatic example of an open quantum system, as it is one of the few exactly solvable open systems \cite{HuPazZhang1992,KarrleinGrabert97,Paz94,GrabertWeiss84}. 
In this section, we present a simple way how to identify the exact master equation for the damped harmonic oscillator as master equation of pseudo-Lindblad form. 
This highlights the relevance of non-GKSL master equation for accurate open system dynamics. 
Furthermore, applying the symmetry transformations of \cref{sec:symmetry_pseudo_Lind}, we make the connection to the approximative equation of Brownian motion. 	
The total system-bath Hamiltonian is given by
\begin{align}
	\hat{H}_\mathrm{tot} =& \frac{\hat{p}^2}{2M} + \frac{M \Omega^2}{2} \hat{q}^2 \notag \\ & + \sum_k^\infty \Bigg[ \frac{\hat{p}_k^2}{2m_k} + \frac{m_k \omega_{k}^2}{2} \Big(\hat{q}_k - \frac{c_k}{m_k \omega_k^2} \hat{q} \Big)^2 \Bigg],
	\label{eq:Htot}
\end{align}
with displacement $\hat{q}$ and momentum $\hat{p}$ of the central oscillator. 
The coupling between system and bath is of the form 
\begin{align}
	\hat{H}_\mathrm{SB} = \hat{q} \otimes \hat{B} ,
\end{align}
with bath operator $\hat{B}=\sum_k^\infty -c_k \hat{q}_k$, where the coefficients $c_k$ are spring constants that de\-ter\-mine the coupling strength between the individual bath modes and the system. 
Yet, there is another term arising from the total system-bath Hamiltonian
\begin{align}
	H_\mathrm{RN} = \sum_k^\infty  \frac{c_k^2}{2 m_k \omega_k^2}\,  \hat{q}^2 = \int_0^\infty \frac{J(\omega)}{\omega} \frac{d\omega}{\pi}\, q^2
\end{align}
with spectral density, $J(\omega) =  \pi \sum_k \frac{c_k^2}{2m_k\omega_k} \delta(\omega-\omega_k)$. 
The additional term in the Hamiltonian is a potential renormalization and cancels the frequency shift caused by the reorganization energy $G_\mathrm{RN}'' = - \int_0^\infty \frac{J(\omega)}{\omega} \frac{d\omega}{\pi}$ [\cref{sec:temp_dep_coupling_dens}].

In the high-temperature regime, the reduced dynamics of the damped harmonic oscillator is described by the equation of Brownian motion, which is a GKSL master equation \cite{breuerpetruccione}. 
In this section, we make the connection between the exact Hu-Paz-Zhang master equation and the pseudo-Lindblad equation, which depends on the parameters $\lambda$ and $\varphi$ as they were discussed in the previous sections. 

The Hu-Paz-Zhang master equation \cite{HuPazZhang1992,KarrleinGrabert97}, also known as Caldeira-Leggett master equation \cite{CaldeiraLegget83}, reads
\begin{align}
	\begin{aligned}
		\partial_t \varrho(t) =&  -\i \bigg[\frac{\hat{p}^2}{2M} + \frac{M \gamma_q(t)}{2} \hat{q}^2, \varrho(t)\bigg]  \\ & - M^2 D_p(t) [q,[q,\varrho(t)]] - \frac{\i}{2} \gamma_p(t) [\hat{q},\{\hat{p},\varrho(t)\}]  \\
		&+MD_q(t) [\hat{q},[\hat{p},\varrho(t)]] . \label{eq:HuPazZhang}
	\end{aligned}
\end{align}
with time-dependent coefficients $\gamma_q(t)$, $\gamma_p(t)$, $D_q(t)$, $D_p(t)$, which are specified, e.g., in Ref.~\cite{KarrleinGrabert97}. We do not explicitly reproduce the expressions for the coefficients here, since their precise form is lengthy and not relevant in the following. 
The potential renormalization $\gamma_q(t)$ and the dissipation rate $\gamma_p(t)$ follow from the classical equation of motion for the damped harmonic oscillator \cite{GrabertWeiss84}.
The diffusion coefficients $D_q(t)$ and $D_p(t)$ follow from fluctuating forces in the bath and are temperature dependent \cite{GrabertWeiss84}. 
The diffusion coefficients $D_q(t)$, $D_p(t)$ are related to the $\gamma_q(t)$, $\gamma_p(t)$ via the quantum fluctuation-dissipation theorem \cite{grabertQuantumBrownianMotion1988}. 

Algebraically, the Hu-Paz-Zhang master equation is quadratic in the operators $q$ and $p$, which are the displacement operator and its canonically associated momentum $p$. 
In addition, the master equation is hermiticity- and trace-preserving and, thus, it can be brought to Redfield form.
Let us now bring \cref{eq:HuPazZhang} into the equivalent form
\begin{align}
	\partial_t \varrho(t)&= -\i \bigg[ \frac{\hat{p}^2}{2M} +\frac{M\gamma_{q}(t)}{2} \hat{q}^2,\  \varrho(t)\bigg] \notag \\
	&+ \Bigg( \hat{\mathbb{S}}^\mathrm{HPZ} \varrho(t)\, \hat q - q\, \hat{\mathbb{S}}^\mathrm{HPZ} \varrho(t) + \mathrm{H.c.} \Bigg),
	\label{eq:HPZasRedfield}
\end{align}
with
\begin{align}
	\hat{\mathbb{S}}^\mathrm{HPZ} = M^2D_p(t)\, \hat{q} + \left(\i \frac{\gamma_p(t)}{2} - MD_q(t)\right) \hat{p}.
	\label{eq:convolution_HPZ}
\end{align}
Since the coupling operator is given by $S=q$, we can now see that the Hu-Paz-Zhang equation as written in \cref{eq:HPZasRedfield} has exactly the form of the Redfield equation, except for the fact that $\mathbb S^\mathrm{HPZ}$ is not given by the convoluted coupling operator [\cref{eq:convolution_operator}], but rather by \cref{eq:convolution_HPZ}. 

Since we have seen that the Hu-Paz-Zhang master equation is of Redfield form, we can now express it also in pseudo-Lindblad form by using the results of the previous sections. 
The Lamb-shift Hamiltonian reads 
\begin{align}
	\hat H^{\mathrm{HPZ}}_\mathrm{LS} &=\frac{\hat q \mathbb{S}^{\mathrm{HPZ}} - \mathbb{S}^{\mathrm{HPZ}\dagger} q}{2\i}=  \frac{\gamma_p(t)}{4} \big\{ \hat q, \hat p\big\} - \frac{MD_q(t)}{2\i} [q,p],  
	\label{eq:HPZlambshift}
\end{align}
where due to $[ q, p]=\i \mathds{1}$, the diffusion parameter $D_q(t)$ gives a constant contribution and might be absorbed in an overall energy shift. 
The jump operators are given by \cref{eq:jumpoperators},
\begin{align}
	A^{\mathrm{HPZ}}_\sigma = \frac{ 1/\lambda }{\sqrt{2\cos \varphi}}  &\Big[ \left(\sigma \lambda^2 e^{-\i \sigma \varphi} +  M^2 D_p(t) \right) q  \notag \\
	&+ \left(\i \frac{\gamma_p(t)}{2} - MD_q(t)\right) \hat{p}  \Big],
\end{align}
where the overall phase factor $e^{\mathrm i \sigma \frac{\varphi}{2}}$ is omitted.

In the following, we apply the optimization procedure to minimize the negative contribution.
In principle, the Hilbert space of the harmonic oscillator is infinite dimensional, such that the norm of the displacement and momentum are not well-defined. 
However, the state space can be truncated at an energy that is large compared to both the initial energy of the system and the temperature of the environment, which is a standard approach in numerical implementations.
As a consequence of the truncation, the commutator of operators is always traceless (as follows immediately from cyclic permutations under the trace $\tr\{AB-BA\}=\tr\{AB-AB\}=0$).
While this breaks the commutation relation of bosonic operators, this artifact of the approximation originates only from the action of the operators on states at the truncation energy, whose population should be negligible.
Namely, in a finite dimensional Hilbert space with cutoff $D$, the product of the bosonic operators in the energy basis is, 
\begin{align}(a a^\dagger)_{nm}=\left( \begin{matrix}
	1 &  0 &\cdots & 0 &0\\
	0 & 2 &\cdots &0&0\\
	\vdots &\vdots    & \ddots & \vdots &\vdots  \\
	0 &  0 &\cdots & D-1 &0\\
	0 & 0 &\cdots&0&0
\end{matrix} \right) ,
\end{align}
and
\begin{align}(a^\dagger a)_{nm}=\left( \begin{matrix}
	0&  0 &\cdots & 0 &0\\
	0 & 1 &\cdots &0&0\\
	\vdots &\vdots    & \ddots & \vdots &\vdots  \\
	0 &  0 &\cdots & D-2 &0\\
	0 & 0 &\cdots&0&D-1
\end{matrix} \right) ,
\end{align}
so that $[a,a^\dagger]\ne1$, i.e.,
\begin{align}([a, a^\dag])_{nm}=\left( \begin{matrix}
	1&  0 &\cdots & 0 &0\\
	0 & 1 &\cdots &0&0\\
	\vdots &\vdots    & \ddots & \vdots &\vdots  \\
	0 &  0 &\cdots & 1 &0\\
	0 & 0 &\cdots&0&1-D
\end{matrix} \right)  \ne \delta_{nm}, 
\end{align}
giving $\tr{[a,a^\dag]}=0$.

For the norm of the jump operators, one then obtains
\begin{align}
	\begin{aligned}
		\lVert {A^\mathrm{HPZ}}^\sigma\rVert^2 &= \frac{1/\lambda^2}{2\cos\varphi} \Big[ \notag \\
		&\big( \lambda^4 + \sigma \lambda^2 M^2 D_p(t) 2\cos\varphi + M^4 D_p(t)^2 \big) \lVert q\rVert^2  \\ 
		&+ \left(\frac{\gamma_p(t)^2}{4} +M^2 D_q(t)^2 \right) \lVert p\rVert^2 \Big] ,
	\end{aligned}
\end{align}
where the mixing term, involving $\tr(qp) = (\mathrm{i}/2)\tr ([a,a^\dagger] )= 0$, vanishes in a finite dimensional state space. 
The latter can be seen from the relations $q = \sqrt{1/2M\Omega} (a^\dagger + a)$ and $p=\mathrm{i} \sqrt{M\Omega/2} (a^\dagger - a)$, which also yield $\lVert p \Vert^2 = M^2\Omega^2 \lVert q\rVert^2$.

The weight difference of the Hu-Paz-Zhang master equation is given by [\cref{eq:net_weight}]
\begin{align}
	\lVert A^\mathrm{HPZ}_+\rVert^2 - \lVert A^\mathrm{HPZ}_-\rVert^2 =2M^2 D_p(t) \lVert q\rVert^2,
\end{align}
and is invariant under the symmetry parameters.

The optimal rescaling $\lambda^\mathrm{HPZ}$ and complex phase $\varphi^\mathrm{HPZ}$, which minimize the negative contribution, are readily obtained from \cref{eq:optimal_lambda,eq:optimal_phi},
\begin{align}
	{\lambda^\mathrm{HPZ}(t)}^2 &
	= M^2 D_p(t)   \sqrt{1 +  \frac{\Omega^2 \gamma_p(t)^2}{4M^2 D_p(t)^2} +  \Omega^2 \frac{D_q(t)^2}{D_p(t)^2}}, \\
	\sin\varphi^\mathrm{HPZ} &\propto \mathrm{Im} \tr( \mathbb S^\mathrm{HPZ} S) = 0 ,
\end{align}
for which we obtain
\begin{align}
	\lVert A^\mathrm{HPZ}_\sigma\rVert^2 & = M^2 D_p(t) \notag \\
	& \left( \sigma +    \sqrt{1 +  \frac{\Omega^2 \gamma_p(t)^2}{4M^2 D_p(t)^2} +  \Omega^2 \frac{D_q(t)^2}{D_p(t)^2}} \right) \lVert q\rVert^2 .
\end{align}

We now make the connection to the GKSL master equation for Brownian motion. In the asymptotic limit with static coefficients, in the high-temperature regime, for weak coupling and large cutoff frequency, the Hu-Paz-Zhang master equation is defined by 
\begin{align}
		\gamma_q &\simeq \Omega^2,  	\label{eq:brownian_motion_gammaq}\\ 
		\gamma_p &\simeq \gamma,   	\label{eq:brownian_motion}\\
		D_q &\simeq0, 	\label{eq:brownian_motion_Dq} \\ 
		D_p &\simeq \frac{\gamma}{M\beta} ,	\label{eq:brownian_motion_Dp}
\end{align}
with coupling rate $\gamma$ and inverse bath temperature $\beta$. 
This is often referred to as quantum Brownian motion or the classical limit in Ref.~\cite{KarrleinGrabert97}. 
With these coefficients, the optimization yields
\begin{align}
		A^{\mathrm{HPZ}}_+&\simeq \sqrt{\frac{\gamma}{2}} \Bigg[ \sqrt{\frac{4M}{\beta}} q + \i \sqrt{\frac{\beta}{4M}} p \Bigg] , \\
		A^{\mathrm{HPZ}}_- &\simeq  \i \sqrt{\frac{\gamma}{2}}  \sqrt{\frac{\beta}{4M}} p,
\end{align}
where the negative contribution has a small weight and can be truncated to arrive at the result of Ref.~\cite{breuerpetruccione}.

We thus provide a generalization to the result of Ref.~\cite{breuerpetruccione}  for arbitrary $\gamma_p(t)$, $D_q(t)$, $D_p(t)$ and have shown that it corresponds to the truncation of the negative contribution after following the optimization procedure.
Beyond the limit of Brownian motion, we provide a better truncation. 
Exploring the accuracy of this truncation is an interesting question for future research.

\section{conclusion}
Often quantum master equations are used to model the dynamics of open quantum systems. 
Beyond ultraweak coupling these master equation are typically not of GKSL form, but instead of pseudo-Lindblad form, which allow for negative relaxation weights.
However, the representation of the pseudo-Lindblad equation is not unique. 
Here, we have used symmetry transformations to systematically construct all possible representations of a generic pseudo-Lindblad equation with two jump operators. 
Then, we showed that this scenario captures two important master equations, the Redfield equation and the Hu-Paz-Zhang master equation. By applying the symmetry transformations to these two examples, we can analytically find the respective optimal representation, which minimizes the negative dissipator. This minimization is optimal both for the approximation given by the truncation of the negative term (so that the resulting master equation is of GKSL form \cite{TBecker2021}) as well as for the efficiency of a recently proposed quantum trajectory unraveling of pseudo-Lindblad equations \cite{beckerQuantumTrajectoriesTimeLocal2023}. 
Thus, the results provided here are of immediate use for the efficient simulation of open quantum systems beyond the limit of ultraweak coupling.

\begin{acknowledgments}
	This research was funded by the Deutsche Forschungsgemeinschaft (DFG) via the Research Unit FOR 2414 under the Project No. 277974659. 
	We thank Isaac Tesfaye and Nathan Harshman for helpful discussions.
\end{acknowledgments}

\bibliography{zotero_library}

\appendix

\section{Microscopic description}
\label{sec:Redfield_eq}
For open quantum systems that are coupled to their environment, an effective description for the reduced dynamics of the system is of great interest. For ultraweak coupling between system and bath, GKSL master equations yield an adequate approximation. Often they are used to describe relaxation, dissipation and decoherence on a phenomenological level  \cite{SDiehlPZoller2008,DPoletti2020,MRehMGaerttner2021,SDenisov2019}. However away from equilibrium, the properties of open quantum systems depend on the details of their environment. A microscopic derivation of a master equation is therefore crucial. 

The microscopic Hamiltonian of the total system-bath compound is denoted as $H_\mathrm{tot} = H_\mathrm{S} + H_\mathrm{SB} + H_\mathrm{B}$
with $H_\mathrm{S}$ and $H_\mathrm{B}$ acting solely on the system and bath, respectively. 
System and bath are coupled via the interaction Hamiltonian $H_\mathrm{SB}$ acting on the joint Hilbert space of system and bath. 
Without loss of generality, the interaction can be written in the canonical form, e.g., by doing a Schmidt decomposition,
\begin{align}
	H_\mathrm{SB} = \sum_\alpha (S_\alpha \otimes B_\alpha^\dagger + S_\alpha^\dagger \otimes B_\alpha),
\end{align}
with $S_\alpha$ referred to as coupling operator solely acting on the system and $B_\alpha$ being some bath operator. 
It is convenient to choose the dimensions of both S and B equal, i.e., the square root of a frequency.
Thereby, multiple coupling operators $S_\alpha$ for $\alpha = 1,2,\dots$ are considered, only, for systems driven out of equilibrium by multiple reservoirs, for which we write $H_\mathrm{B} = \sum_\alpha {H_\mathrm{B}}_\alpha$. 
To keep the notation simple, here we consider one coupling channel with Hermitian $S$ in $H_\mathrm{SB} = S \otimes (B^\dagger + B)$. 
Generalizations are easily found by reintroducing the summation over $\alpha$. 

We are interested in the effective dynamics of the reduced state of the system by tracing over the bath degrees of freedom, i.e., $\varrho(t) = \tr_\mathrm{B} \varrho_\mathrm{tot}(t)$. In a perturbative treatment up to second order in the system-bath coupling, this is widely known as Born-Markov theory, e.g., in Ref.~\cite{breuerpetruccione,weissQuantumDissipativeSystems2012}. 
One obtains the time-local quantum master equation 
\begin{align}
	\partial_t \varrho(t) =- \i [H_\mathrm{S}, \varrho(t)] + \mathcal R^{(2)}[ \varrho(t)] , 
\end{align}
with Redfield superoperator 
\begin{alignat}{2}
			\mathcal R^{(2)}[\varrho(t)] =  \mathbb S \varrho(t) S + \hat{S}\hat{\rho}(t) \hat{\mathbb{S}}^\dagger  -  \hat{S} \hat{\mathbb{S}} \hat{\rho}(t) -  \hat{\rho}(t)\hat{\mathbb{S}}^\dagger\hat{S}  , \label{eq:redfield}
\end{alignat}
for which we define the (time-dependent) convolution operator 
\begin{align} 
	\hat{\mathbb{S}} = \int_0^{t} C(\tau)\, \mathbf{S}(-\tau)\ d\tau , 
\end{align}
which possesses the same dimension as $S$, and bath correlation 
 \begin{align} 
 	C(\tau)=\expval{ B(\tau) B^\dagger} +\expval{ B^\dagger(\tau) B} ,
 \end{align} 
with $ B(\tau) = e^{\i H_\mathrm{B} \tau} B   e^{-\i H_\mathrm{B} \tau}$. 

Further, one completes the latter terms in \cref{eq:redfield} to a commutator and anti-commutator $- S \mathbb{S} \varrho(t) - \varrho(t) \mathbb{S}^\dagger S = -\i [ (S\mathbb{S} - \mathbb{S}^\dagger S)/(2 \i), \varrho(t)] - (1/2) \{S \mathbb{S}+\mathbb{S}^\dagger S, \varrho(t) \}$ to arrive at
\begin{align}
	\mathcal R^{(2)}[\varrho(t)] &= - \i [H_\mathrm{LS} , \varrho(t) ] + \mathcal D^\mathrm{Red}[\varrho(t)] , 
\end{align}
with Lamb-shift Hamiltonian
\begin{align}
	H_\mathrm{LS} = \frac{S\mathbb{S} - \mathbb{S}^\dagger S}{2 \i} . 
\end{align}
The dissipative part has the familiar form 
\begin{align}
	\mathcal D^\mathrm{Red}[\varrho(t) ] = \sum_{i,j=1}^2 \sigma^x_{ij} \Bigg[ S_i \varrho(t) S_j^\dagger - \frac{1}{2} \Big\{ S_j^\dagger S_i, \varrho(t) \Big\}\Bigg]  , \label{eq:redfield_dissipator} 
\end{align}
with Kossakowski matrix
\begin{align}
	\sigma^x &= \left(\begin{array}{rr}
		0 & \quad 1\\
		1 & \quad 0 \\
	\end{array}\right) ,
\end{align}
given by the Pauli-$x$-matrix and jump operators $S_1\equiv S$, $S_2 \equiv \mathbb S$. 

Whereas the Lamb-shift is well known for the quantum-optical master equation, i.e., within the rotating-wave approximation, the identification made here for the Redfield equation can be found, e.g., in Refs.~\cite{farinaOpenquantumsystemDynamicsRecovering2019,TimofeevTrushechkin2022}. 
The Redfield Lamb-shift can be used as the basis for defining thermodynamic properties, such as heat and work beyond ultraweak coupling \cite{CollaBreuer2022}. To this end, we remark that the operators in the dissipative part may be transformed by redefining the Lamb-shift. 
In reference \cite{CollaBreuer2022} a unique choice of the Lamb-shift is proposed by minimizing the dissipator. 
In this work, we do not make use of this transformation but solely focus on the dissipative part defined above.

\section{Unitary diagonalization of the Redfield equation in the system's eigenbasis}
\label{sec:analytical_diagonalization}

We consider the Redfield equation represented in the eigenbasis of the system, i.e., $H_\mathrm{S} \ket{q} = E_q \ket{q}$
The convolution operator takes the form $\mathbb S_t = \sum_{qk} G_t(\Delta_{qk}) S_{qk} L_{qk}$, with jump operators $\hat{L}_{qk}=\ketbra{q}{k}$ and frequencies $\Delta_{qk}=E_q - E_k$. 
Thereby, 
\begin{align}
	G_t(\Delta) \equiv \int_0^t e^{-\i \Delta \tau} C(\tau) \mathrm{d}\tau ,
\end{align}
defines the temperature-dependent coupling density [\cref{sec:temp_dep_coupling_dens}].
In the asymptotic limit $\lim_{t\to \infty} G_t(\Delta)=G(\Delta)$ it is the half-sided Fourier transform of the bath correlation.  

Because the Redfield equation is quadratic in the coupling operator the summation runs over four indices $q,k$ and $q',k'$ for the two frequencies $\Delta_{qk}$ and $\Delta_{q'k'}$ and can be brought to the form
\begin{widetext}
	\begin{align}
		\begin{split}
			\mathcal{R}^{(2)}[\varrho(t)]  =&  - \i \left[ \sum_{qk, q'k'} S_{qk}S_{k'q'} \frac{G(\Delta_{qk}) - G(\Delta_{q'k'})^*}{2\i} L_{q'k'}^\dagger L_{qk}  , \varrho(t) \right]  \\&+ \sum\limits_{qk, q'k'} S_{qk} S_{k'q'}  \left(G(\Delta_{qk}) + G(\Delta_{q'k'})^*\right) \Big[ \hat{L}_{qk} \varrho(t) \hat{L}_{q'k'}^\dagger-  \frac{1}{2} \Big\{ \hat{L}_{q'k'}^\dagger \hat{L}_{qk} ,\varrho(t)  \Big\} \Big],
		\end{split}
		\label{eq:redfield_in_components}
	\end{align}
\end{widetext}
which is the full Redfield superoperator and includes the Lamb-shift Hamiltonian $H_\mathrm{LS}$ in the first line and the dissipator $\mathcal D^\mathrm{Red}$ in the second line. 

In the eigenbasis, the Redfield equation in (\ref{eq:redfield_in_components}) resembles the GKSL structure, except it is not diagonal in the jump operators $S_{qk} L_{qk}$, with $L_{qk} = \dyad{q}{k}$. 
We obtain a diagonal representation by solving the eigenvalue problem of the $D^2$-dimensional, Hermitian coefficient matrix $M$ with entries $M_{(qk),(q'k')} = G(\Delta_{qk}) + G(\Delta_{q'k'})^* $. 
The eigenvalue problem reads
\begin{align}
	\sum_{q'k'} \big(G(\Delta_{qk}) + G(\Delta_{q'k'})^*\big)\, v^{\sigma}_{q'k'} = g_\sigma \, v^\sigma_{qk} ,
\end{align}
with real eigenvalues $g_\sigma$ and eigenvectors $v^\sigma = (v^\sigma_{qk})$. 

A numerical diagonalization for large systems is very demanding and lacks for a detailed understanding. Therefore, as an unprecedented result, we analytically solve the eigenvalue problem using simple algebraic methods. First, we find a closed quadratic equation for the non-zero eigenvalues $g_\sigma$, and secondly use this to construct the eigenvectors $v^\sigma$. 

Thereby, it is useful to introduce averages of the form,
\begin{align}
	\expval{x} \equiv \frac{1}{D^2} \sum_{qk} x_{qk},
\end{align}
such as $\expval{v^\sigma} \equiv \frac{1}{D^2} \sum_{qk} v^\sigma_{qk}$, $\expval{G^*v^\sigma} \equiv (1/D^2) \sum_{qk} G(\Delta_{qk})^* v^\sigma_{qk}$, etc., and variance $V[x] \equiv  \expval{x^2} - \expval{x}^2$. 
As starting point, we formally solve the eigenvalue problem for $v^\sigma_{qk}$,  assuming $g_\sigma\ne0$,
\begin{align}
	v^\sigma_{qk} = \frac{D^2}{g_\sigma} \left( G(\Delta_{qk})\, \expval{v^\sigma} + \expval{G^*v^\sigma} \right),
	\label{eq:eigenvalue_equation_for_Kossak}
\end{align}
and find it to be affine-linear in $G(\Delta_{qk})$. In the following, on the right-hand side, we get rid of the implicit dependence on the eigenvector $v^\sigma$ by taking appropriate averages. Namely, by taking the average of \cref{eq:eigenvalue_equation_for_Kossak} we find the relation $\expval{v^\sigma} = \frac{D^2}{g_\sigma} \left( \expval{G}\, \expval{v^\sigma} + \expval{G^*v^\sigma}\right)$.

Firstly, for the eigenvalue $g_\sigma$, we eliminate $\expval{v^\sigma} = \expval{G^*v^\sigma}/(\frac{g_\sigma}{D^2} - \expval{G})$ and arrive at,
\begin{align}
	v^\sigma_{qk} = \frac{D^2}{g_\sigma} \expval{G^*v^\sigma} \left( \frac{G(\Delta_{qk})}{\frac{g_\sigma}{D^2} - \expval{G}} + 1 \right).
\end{align}
Here, in order to cancel $\expval{G^*v^\sigma}$ we multiply by $G(\Delta_{qk})^*$ and, yet again, take the average. It follows, $1 =  \frac{D^2}{g_\sigma} \big( \frac{\expval{|G|^2}}{\frac{g_\sigma}{D^2} - \expval{G}} + \expval{G^*} \big)$.
This is a quadratic equation for $g_\sigma$, $0=(\frac{g_\sigma}{D^2} - \expval{G^*} ) (\frac{g_\sigma}{D^2} - \expval{G} ) - \expval{|G|^2}$, which has the solution $\frac{g_\sigma}{D^2} = \expval{G} + \expval{G^*} \pm \sqrt{( \expval{G} + \expval{G^*})^2 + \expval{|G|^2} - |\expval G|^2 }$. With real and imaginary part, i.e., $G = G' + \mathrm i G''$, it follows $\frac{g_\sigma}{D^2}= \expval{G'} \pm \sqrt{\expval{G'}^2 + \expval{G'^2} - \expval{G'}^2 + \expval{G''^2} - \expval{G''}^2}$, which can be written as
\begin{align}
	\frac{g_\sigma}{D^2} = \expval{G'} \left(1 + \sigma \sqrt{1 + \frac{V[G'] +V[G'']}{\expval{G'}^2}} \right),
	\label{eq:eigenvalue_Kossakowski}
\end{align}
with $\sigma=\pm1$.

Secondly, for the eigenvectors $v^\sigma$, we start again from \cref{eq:eigenvalue_equation_for_Kossak} and eliminate $\expval{G^*v^\sigma} = \expval{v^\sigma} (\frac{g_\sigma}{D^2} - \expval{G})$,
\begin{align}
	v^\sigma_{qk} = \frac{D^2}{g_\sigma} \expval{v^\sigma}\left( G(\Delta_{qk}) + \sigma \sqrt{ \expval{|G|^2}} e^{-\mathrm i \sigma \varphi_0} \right),
\end{align}
where the second term is given by the complex quantity
\begin{align}
	\begin{split}
		\frac{g_\sigma}{D^2} - \expval{G} 
		&= \sigma \sqrt{ \expval{G'}^2 + V[G'] + V[G'']} - \mathrm i \expval{G''} ,\\
		&=\sigma \sqrt{\expval{|G|^2} - \expval{G''}^2} - \mathrm i \expval{G''} ,\\
		&\equiv \sigma \sqrt{\expval{|G|^2}} e^{-\mathrm i \sigma \varphi_0} ,
	\end{split}
\end{align}
for which, in the last line, $\varphi_0$ defines the complex phase, which can be interpreted geometrically in the figure below. 
For the complex phase, it is $\tan \varphi_0=  \frac{\overline{G''}}{\sqrt{\overline{|G|^2} - \overline{G''}^2}}$, $\sin \varphi_0=\frac{\overline{G''}}{\sqrt{\overline{|G|^2}}}$, $\cos \varphi_0=\frac{\sqrt{\overline{|G|^2} - \overline{G''}^2}}{\sqrt{\overline{|G|^2}}}$ with $\varphi_0 \in [0,\frac{\pi}{2})$.

\begin{figure}[b!]
	\centering
	\begin{tikzpicture}[
		my angle/.style = {draw,
			angle radius=15mm, 
			angle eccentricity=1.1, 
			left, inner sep=2.5mm,
			font=\normalsize} 
		]
		\draw[thick]   (0,0) coordinate[label=below:] (a) --
		(5,0) coordinate[label=below:] (c) --
		(5,-2) coordinate[label=above:] (b) -- cycle;
		\pic[my angle, "$\varphi_0$"] {angle = b--a--c};
		\node[left] at (4.5,0.5) {$\mathrm{Re} z = \sqrt{\expval{|G|^2} - \expval{G''}^2}$};
		\node[left] at (7.5,-1) {$\mathrm{Im} z =- \expval{G''}$};
		\node[left] at (3.2,-1.7) {$|z| = \sqrt{\expval{|G|^2}}$};
	\end{tikzpicture}
	\caption[Complex quantity $z=\frac{g_\sigma}{D^2} - \expval{G} $ visualized in the complex plane.]{Complex quantity $z=\frac{g_\sigma}{D^2} - \overline{G} $ visualized in the complex plane.}
	\label{fig:complex_quantity}
\end{figure}
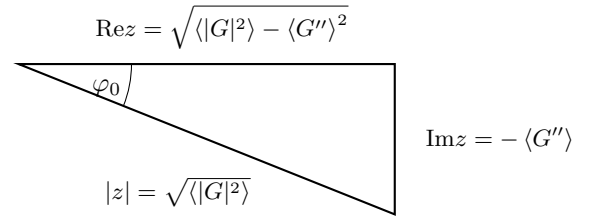

As a final step, the vector $v^\sigma$ is normalized. 
The squared norm of the eigenvector is given by 
\begin{align} 
 |v^\sigma|^2 &= \sum_{q'k'} |v^\sigma_{q'k'}|^2 , \notag \\
&= \frac{D^4}{g_\sigma^2} \lvert {\expval v^\sigma} \rvert^2 \sqrt{\expval{ |G|^2}} 2 (\sqrt{\expval{|G|^2}} + \sigma \mathrm{Re}[ e^{\mathrm i \sigma \varphi_0} \expval{G}] ) 
, \notag \\ 
\begin{split}
&= \frac{D^4}{g_\sigma^2} \lvert {\expval v^\sigma} \rvert^2 \sqrt{\expval{ |G|^2}} 2 \cos\varphi_0  \\ 
& \qquad \Big( \frac{\sqrt{\expval{|G|^2}}}{\cos\varphi_0} + \sigma \expval{G'} - \tan\varphi_0\, \expval{G''} \Big)  .
\end{split}
\end{align} 
Using the expressions for the complex phase [see \cref{fig:complex_quantity}] the latter term reduces to $ \frac{\sqrt{\expval{|G|^2}}}{\cos\varphi_0} + \sigma \expval{G'} - \tan\varphi_0 \, \expval{G''} = \sqrt{\expval{G'^2} + V[G'] + V[G'']} + \sigma \expval{G'} = \frac{\sigma g_\sigma}{D^2} =  \frac{|g_\sigma|}{D^2}  $. 
We arrive at the normalized eigenvector
\begin{align}
	\frac{v^\sigma_{qk}}{|v^\sigma|} = \frac{1}{\sqrt{2 |g_\sigma| \cos\varphi_0}} \left( \frac{1}{\lambda_0} e^{\mathrm i \sigma \frac{\varphi_0}{2}} G(\Delta_{qk}) + \sigma \lambda_0 e^{-\mathrm i \sigma \frac{\varphi_0}{2}} \right),
\end{align}
for which we neglect the irrelevant phase factor $\frac{\expval{v^\sigma}}{|\expval{v^\sigma}|} e^{-\mathrm i \frac{\varphi_0}{2}}$ and introduce the rescaling parameter $\lambda_0^2 \equiv \sqrt{\expval{|G|^2}}$ to match with the notation of \cref{eq:jumpoperators}.

In summary, by diagonalization of the coefficient matrix,
\begin{align}
	G(\Delta_{qk}) + G(\Delta_{q'k'})^* = \sum_{\sigma=+1,-1} g_\sigma \frac{v^\sigma_{qk}}{\lvert v^\sigma \rvert} \frac{{(v^\sigma_{q'k'})}^*}{\lvert v^\sigma \rvert},
\end{align}
the Redfield equation (\ref{eq:redfield_in_components}) takes a pseudo-Lindblad form with dissipator 
\begin{align} 
	\mathcal{D}^\mathrm{Red}[\hat{\rho}]  =  \sum_{\sigma} \sigma \allowbreak \Big[\hat{A}^0_\sigma \hat{\rho} \hat{A}_\sigma^{0\dagger} - \frac{1}{2} \Big\{ \hat{A}_\sigma^{0\dagger} \hat{A}^0_\sigma , \hat{\rho}  \Big\} \Big], 
\end{align}	
and jump operators,
\begin{align}
	A_\sigma^0
	&= \sum\limits_{qk} \sqrt{|g_\sigma|} \frac{v^\sigma_{qk}}{\lvert v^\sigma \rvert} S_{qk} \dyad{q}{k} , \notag \\
	&= \frac{1}{\sqrt{2\cos \varphi_0}}  \left( \frac{1}{\lambda_0} e^{\mathrm i \sigma \frac{\varphi_0}{2}} \mathbb S + \sigma \lambda_0 e^{-\mathrm i \sigma \frac{\varphi_0}{2}}  S \right).
\end{align}
For a discussion of the eigenvalues $g_\sigma$ and jump operators $A^\sigma$, we refer to \cref{sec:minimize_neg_weight} of the main text. 
Surprisingly, there we find that the parameters $\lambda_0$ and $\varphi_0$ can be chosen arbitrarily

\section{Symmetry transformation for the (non-diagonal) Redfield equation }
\label{sec:symmetry_Redfield}
In the following, the symmetry transformation $\Lambda$ for the Kossakowski matrix $\sigma^x$ of the Redfield equation is constructed, $\sigma^x = \Lambda \sigma^x \Lambda^\dagger$. 

Starting with four arbitrary complex entries in $\Lambda$, 
\begin{align}
	\Lambda = \left(\begin{array}{rr}
		\Lambda_{11} &  \Lambda_{12}\\
		\Lambda_{21} &  \Lambda_{22} \\
	\end{array}\right) ,
\end{align}
the symmetry condition yields the three coupled equations
\begin{widetext}
\begin{align}
	0 &= 2 \mathrm{Re}[ \Lambda_{11} \Lambda_{12}^* ]  = 2 | \Lambda_{11}| |\Lambda_{12}| \cos( \mathrm{arg}(\Lambda_{11} ) - \mathrm{arg}(\Lambda_{12})) ,  \label{eq:pauli_x_1} \\
	0 &= 2 \mathrm{Re}[ \Lambda_{22} \Lambda_{21}^*] = 2 | \Lambda_{22}| |\Lambda_{21}| \cos( \mathrm{arg}(\Lambda_{22} ) - \mathrm{arg}(\Lambda_{21})), \label{eq:pauli_x_2}  \\
	1 &= \Lambda_{11} \Lambda_{22}^* + \Lambda_{12} \Lambda_{21}^* \notag \\ 
	&=\Big( |\Lambda_{11} |\Lambda_{22}| e^{\i (\mathrm{arg}(\Lambda_{11}) - \mathrm{arg}(\Lambda_{12}) + \mathrm{arg}(\Lambda_{21}) - \mathrm{arg}(\Lambda_{22}))} + |\Lambda_{12}| |\Lambda_{21}| \Big) e^{\i (\mathrm{arg}(\Lambda_{12}) - \mathrm{arg}(\Lambda_{21}))} , \label{eq:pauli_x_3} 
\end{align}
\end{widetext}
where $\mathrm{arg}(\Lambda_{ij})$ denotes the complex phase for the matrix element $\Lambda_{ij} = |\Lambda_{ij}| e^{\mathrm i \mathrm{arg}(\Lambda_{ij})}$. 
The conditions above correspond to four real-valued equations that fix four out of eight parameters in $\Lambda$. 

In more detail, from the first and second line one obtains the phase differences $\mathrm{arg}(\Lambda_{11}) - \mathrm{arg}(\Lambda_{12}) = \pi/2 + k\pi$ and $\mathrm{arg}(\Lambda_{21}) - \mathrm{arg}(\Lambda_{22}) = \pi/2 + l \pi$ with $k,l\in\mathbb{Z}$. 
With this, it follows in the third line that the factor $ e^{\i (\mathrm{arg}(\Lambda_{11}) - \mathrm{arg}(\Lambda_{12}) + \mathrm{arg}(\Lambda_{21}) - \mathrm{arg}(\Lambda_{22}))} $ is real valued. 
Consequently, the second factor $e^{\i (\mathrm{arg}(\Lambda_{12}) - \mathrm{arg}(\Lambda_{21}))}$ must also be real, so that one obtains $\mathrm{arg}(\Lambda_{12}) - \mathrm{arg}(\Lambda_{21}) = l'\pi$ with $l'\in\mathbb Z$. Altogether, apart from a global phase, one deduces the diagonal entries to be real and the off-diagonals to be purely imaginary. 
\begin{align}
		\Lambda = e^{\i \alpha} \left(\begin{array}{rr}
		\lambda_{11} &  \i \lambda_{12}\\
		\i \lambda_{21} &  \lambda_{22} \\
	\end{array}\right) ,
\end{align}
with $\lambda_{ij} \in \mathbb R$. 
In this parametrisation, the third condition in \cref{eq:pauli_x_3} reduces to $1 = \lambda_{11} \lambda_{22} + (\i \lambda_{12}) (-\i \lambda_{21}) = \mathrm{det}\Lambda$. 
This condition is met with the parametrisation
\begin{align}
	\Lambda = \frac{e^{\i \alpha}}{\sqrt{\cos{\varphi}}} \left(\begin{array}{rr}
		\lambda \cos(\frac{\varphi+\beta}{2}) &  \i \lambda \sin(\frac{\varphi+\beta}{2})\\
		-\i \frac{1}{\lambda} \sin(\frac{\varphi-\beta}{2}) &  \frac{1}{\lambda} \cos(\frac{\varphi-\beta}{2})\\
	\end{array}\right) .
\end{align}

Together with the unitary transformation,
\begin{align}
	U = \frac{1}{\sqrt 2}  \left(\begin{array}{rr}
		1 &  -1\\
		1 & 1 \\
	\end{array}\right) ,
\end{align}
the jump operators for the pseudo-Lindblad representation follow from the transformation 
\begin{align}
	\Lambda U =
	\begin{split}
		\frac{e^{\i\alpha}}{\sqrt{ 2 \cos \varphi}} 
		\left(\begin{array}{rr}
			\lambda e^{\i \frac{\varphi + \beta}{2}} & -\lambda e^{-\i \frac{\varphi + \beta}{2}} \\
			\frac{1}{\lambda} e^{-\i \frac{\varphi - \beta}{2}} &  \frac{1}{\lambda} e^{\i \frac{\varphi - \beta}{2}}  \\
		\end{array}\right) ,
	\end{split}
\end{align}
and are given by $(A^{(\lambda,\varphi)}_+, A^{(\lambda,\varphi)}_-) = (S,\mathbb S)\, \Lambda U$ in \cref{eq:jumpoperators}.

\section{Temperature-dependent coupling density $G(\Delta)$}
\label{sec:temp_dep_coupling_dens}
Typically, for open quantum systems the bath is assumed to remain in a thermal state. The bath excitations, the bath correlations respectively, decay on a time scale that is fast as compared to the dynamics of the system. In the Markovian limit all excitations leaving the system decay fast enough, such that they do not act back on the system at a later time.
In this regime, the dynamics of the system is described by a time local generator. 
In a perturbative treatment up to second order in the interaction Hamiltonian $\hat{H}_\mathrm{SB}$, the Redfield master equation (\ref{eq:redfield}) is obtained.
Therein, the bath correlation $C(\tau)$ appears in a convolution integral with the Heisenberg operator ${\hat{S}}(-\tau)=e^{-\i \hat{H}_\mathrm{S}\tau}\hat{S}e^{\i \hat{H}_\mathrm{S}\tau}$.

As a first step, the Heisenberg operator is formally represented in the eigenbasis of system, such that the time-dependent convolution can be written as $ \hat{\mathbb{S}}_t = \sum_{qk} G_t(\omega_q - \omega_k) |q\rangle\langle q|\hat{S}|k\rangle\langle k|$, where $\omega_q$ and $\ket{q}$ are the eigenfrequencies and eigenstates of $\hat{H}_\mathrm{S}$. 
We define the temperature dependent coupling density \cite{AEisfeld2012}
\begin{align}
	G_t(\Delta) \equiv \int_0^t e^{-\i \Delta \tau} C(\tau) \mathrm{d}\tau ,
\end{align}
which is also referred to as \emph{spectral noise power}, e.g., in Ref.~\cite{xuPerformancePerturbativeTreatments2023}. 
However, in this work the integration always starts at time $t=0$. Therefore the coupling density is the half-sided Fourier transform and is a complex quantity.
In the asymptotic limit $\lim_{t\to \infty} G_t(\Delta)=G(\Delta)$ this expression becomes the half-sided Fourier transform of the bath correlation.  
By inserting the integral expression $C(t) =  \int_{-\infty}^\infty e^{\i \omega t} J(\omega) n_\beta(\omega) \frac{\mathrm{d} \omega}{\pi}$, with $n_\beta(\omega)=[\exp(\beta \omega) -1]^{-1}$ one arrives at
\begin{align}
	G(\Delta) = J(\Delta)  n_\beta(\Delta) + \i\ \pv{\int_{-\infty}^\infty \frac{J(\omega)  n_\beta(\omega) }{\omega- \Delta} \frac{\mathrm{d} \omega}{\pi}},
	\label{eq:coupling_density}
\end{align}
where the time integration is already performed by making use of the Sokhotskyi-Plemelj formular, 
\begin{align} 
	\int_{0}^{\infty} e^{\i E \tau} \mathrm{d}\tau = \pi \delta(E) + \i \pv{\frac{1}{E} },
\end{align}
which often referred to as Kramers-Kronig relation. 

For second-order master equations the coupling density is to be evaluated at the frequencies $\Delta_{qk}=E_q - E_k$, associated with the energy level splittings, and the quantities $G(\Delta_{qk})$ appear in the tensor elements of the Redfield equation. Even though some literature refers to them as transition rates in general this term might be misleading:
On the one hand, there exist different notations, i.e.~for golden rule type expressions $G(\Delta)=2\pi J(\Delta)n_\beta(\Delta)$ or different definitions of the coupling strength $G(\Delta)\propto \gamma^2$, e.g.~in Ref.~\cite{DWHoneKetzmerick2009}. 
On the other hand, a simple physical interpretation for the coupling density might be possible only for the quantum-optical master equation, for which the diagonal elements of the density matrix obey a Pauli rate equation. Thereby the rate for the transition $\ket{q}\rightarrow \ket{k}$ between energy eigenstates is given by $R_{kq}=2G'(\Delta_{qk}) \lvert\matrixelement{q}{\hat{S}}{k}\rvert^2$ and obeys the detailed-balance condition. 
Also, for the quantum-optical master equation the imaginary part of the bath-correlation function only modifies the coherent dynamics by shifting the eigenenergies. 
Since in this particular case, it only affects the transient dynamics, but not the steady state, it is often neglected. 
However, beyond the secular approximation, the imaginary part of the bath correlation function is of crucial importance. 
In fact, it mainly determines the difference between the quantum optical and Redfield master equation \cite{JThingaPHaenggi2012}. 
In the following, some basic properties of $G''(\Delta)$ are discussed that stem directly from the integral representation. 
By using the antisymmetry of the spectral density $J(\Delta)$, the region of integration can be limited to positive frequencies only, 
\begin{widetext}
\begin{align}
	\begin{aligned}
		G''(\Delta) &= \pv{\int_0^\infty \frac{J(\omega)}{\omega^2-\Delta^2} \big[\omega (n_\beta(\omega) + n_\beta(-\omega))  +  \Delta (n_\beta(\omega) - n_\beta(-\omega))\big] \frac{\mathrm d \omega}{\pi} } \\ 
		&= \underbrace{- \int_0^\infty \frac{J(\omega)}{\omega} \frac{\mathrm d \omega}{\pi} }_{G''_\mathrm{RN}} \underbrace{+ \pv{\int_0^\infty \frac{J(\omega)}{\omega^2 - \Delta^2} \left[ \Delta - \frac{\Delta^2}{\omega} \right] \frac{\mathrm d \omega}{\pi}}}_{G''_0(\Delta)} + \underbrace{2\Delta \pv{\int_0^\infty  \frac{J(\omega) n_\beta(\omega)}{\omega^2-\Delta^2} \frac{\mathrm{d} \omega}{\pi}}}_{G''_\mathrm{th}(\Delta)},
	\end{aligned}
	\label{eq:lambshift_rates}
\end{align}
\end{widetext}
where we used the symmetries of the Bose function $n_\beta(\omega) + n_\beta(-\omega)=-1$ and $n_\beta(\omega) - n_\beta(-\omega) = 2n_\beta(\omega)+ 1$. 
Above, $G''_\mathrm{RN}$ is the reorganization energy, which gives a contribution to the Lamb-shift Hamiltonian \cite{weissQuantumDissipativeSystems2012}. 
One also identifies zero-point fluctuation $G''_0(\Delta)$ and thermal noise $G''_\mathrm{th}(\Delta)$. 
The wording \textit{zero-point fluctuation} and \textit{thermal noise} are vaguely used in the literature and might not agree with the definition given in this paper. 
Though a very similar discussion is given in Refs.~\cite{TAlbash2012,CWGardiner00}. 
Note, the identification of the latter as thermal noise is based on the fact that it vanishes for zero temperature. 
In this limit, the rates to higher excited states are exponentially suppressed, which in the quantum-optical master equation leads to relaxation towards the ground state. 
Formally, for zero temperature in \cref{eq:coupling_density} one replaces  $n_\beta(\omega)= -\Theta(-\omega)$, for which only $G''_\mathrm{RN} + G''_0(\Delta)$ contributes to the imaginary part.

\end{document}